\documentclass[runningheads]{llncs}
\usepackage[T1]{fontenc}
\usepackage{graphicx,verbatim}
\usepackage{booktabs}
\usepackage{amsmath}
\usepackage{amssymb}
\usepackage{multirow}
\usepackage[colorlinks=true]{hyperref}
\hypersetup{
    colorlinks=true,
    linkcolor=blue,
    filecolor=magenta,      
    urlcolor=cyan,
    pdfpagemode=FullScreen,
    }

\begin{document}
\title{Revisiting Integration of Image and Metadata for DICOM Series Classification: Cross-Attention and Dictionary Learning}
\titlerunning{Revisiting Integration of Image and Metadata}
%


\author{Tuan Truong\inst{1} \and
Melanie Dohmen\inst{1} \and Sara Lorio\inst{1} \and
Matthias Lenga \inst{1}}
\authorrunning{Truong et al.}
%
\institute{Bayer AG, Berlin, Germany
}


  
\maketitle              
\begin{abstract}

Automated identification of DICOM image series is essential for large-scale medical image analysis, quality control, protocol harmonization, and reliable downstream processing. However, DICOM series classification remains challenging due to heterogeneous slice content, variable series length, and entirely missing, incomplete or inconsistent DICOM metadata.
We propose an end-to-end multimodal framework for DICOM series classification that jointly models image content and acquisition metadata while explicitly accounting for all these challenges.
(i) Images and metadata are encoded with modality-aware modules and fused using a bi-directional cross-modal attention mechanism. 
(ii) Metadata is processed by a sparse, missingness-aware encoder based on learnable feature dictionaries and value-conditioned modulation. By design, the approach does not require any form of imputation.
(iii) Variability in series length and image data dimensions is handled via a 2.5D visual encoder and attention operating on equidistantly sampled slices. 
We evaluate the proposed approach on the publicly available Duke Liver MRI dataset and a large multi-institutional in-house cohort, assessing both in-domain performance and out-of-domain generalization. Across all evaluation settings, the proposed method consistently outperforms relevant image-only, metadata-only and multimodal 2D/3D baselines. The results demonstrate that explicitly modeling metadata sparsity and cross-modal interactions improves robustness for DICOM series classification. 
\\

Source code: \url{https://github.com/MatthiasLen/img-meta-cls}


\end{abstract}
\section{Introduction}
\begin{figure}[t]
    \centering
    \includegraphics[width=0.95\textwidth, trim={0.5cm 2.5cm 0cm 2.3cm}, clip]{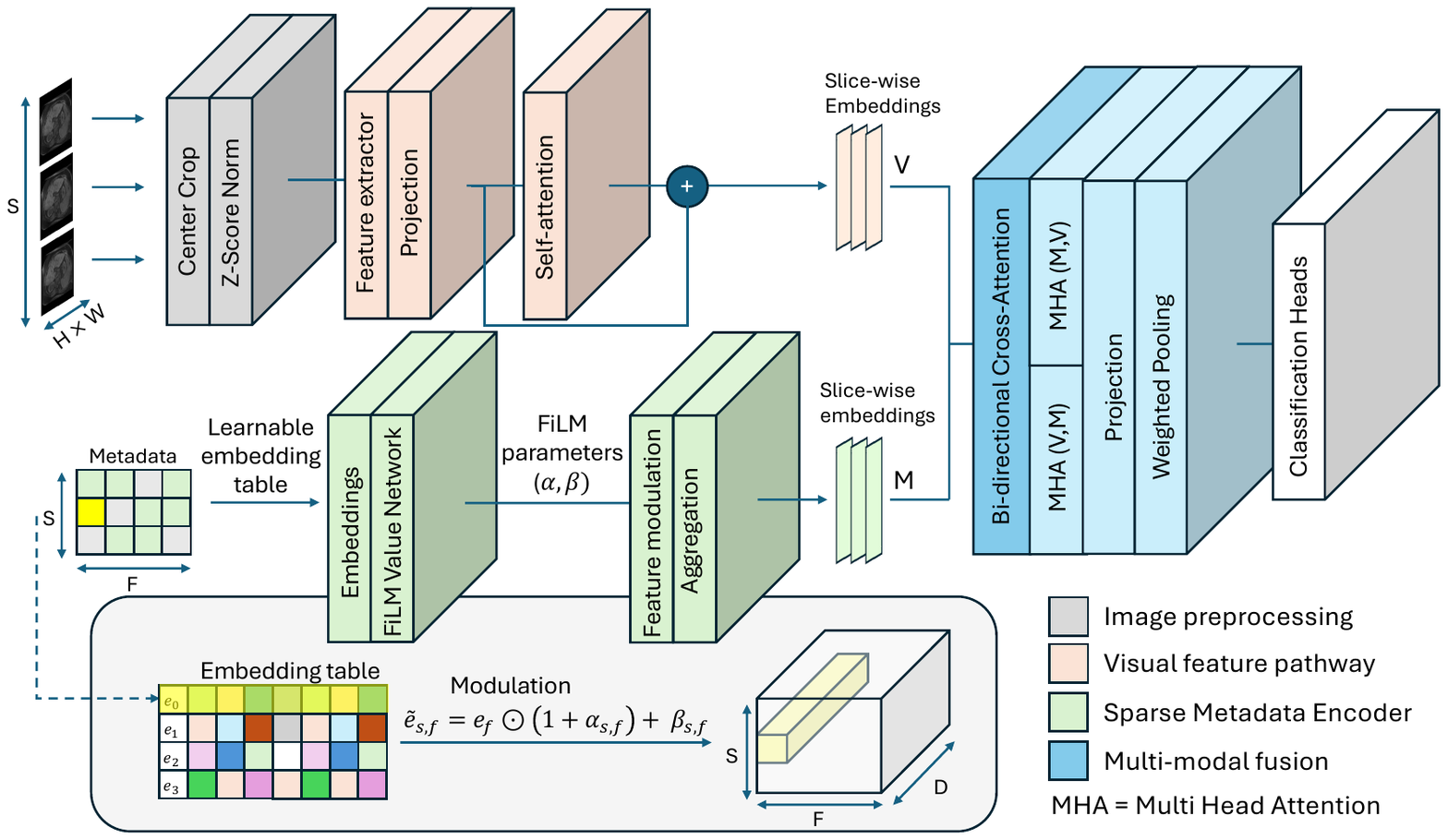}
    \caption{Proposed method: pixel data of $S$ DICOM slices is embedded in visual feature pathway. DICOM metadata is embedded by the Sparse Metadata Encoder. Bi-directional cross-modal attention contextualizes all image and metadata embeddings. Final integration to a series-level representation is done by learnable pooling.}
    \label{fig:architecture}
\end{figure}
%
Medical imaging examinations typically comprise multiple acquisition types, such as different MRI sequence protocols, each providing complementary diagnostic information. Correct identification of DICOM image series is a prerequisite for many downstream tasks, including automated analysis, quality control, dataset curation, or protocol harmonization. Manual identification is time-consuming and error-prone, motivating the development of automated methods for DICOM series classification.
%
Although the DICOM Series Description tag often contains information about sequence type, acquisition plane, or contrast phase, it is inherently unreliable. Metadata fields are vendor-dependent, frequently manually edited, and rarely follow standardized naming conventions.
Recent work has shown that analyzing a wider range of DICOM tags can yield better classification results \cite{gauriau_using_2020,liang_magnetic_2021}. However, metadata-based approaches remain vulnerable to missing, inconsistent, or inaccurate header information. Additionally, certain sequence types, particularly those related to contrast phases, may not be easily inferred from DICOM metadata alone since acquisition parameters may overlap. 
%
Image-based methods, while generally not reliant on metadata quality and availability, face other challenges. Robust series classification approaches require efficiently capturing volumetric context, identifying and aggregating information across slices, handling variations in slice orientation and spacing, and generalizing across different scanners, protocols, and patients. Prior work has explored fine-grained image models and 3D architectures or slice-wise voting \cite{kim_automated_2025-1,kim_automated_2025,yuan_fine-grained_2025,zhu_3d_2022} to effectively leverage visual features in a volumetric context. 

To overcome the limitations of unimodal approaches, multimodal methods combining image data and metadata have recently been explored. Existing solutions often rely on two-stage pipelines that train separate metadata and image classifiers and combine their predictions afterwards \cite{cluceru_improving_2023,miller_automated_2024}. Two-stage approaches prevent joint representation learning and often require prior metadata imputation, which introduces additional sources of error when missingness is substantial. 
In this work, we propose an end-to-end multimodal framework for DICOM series classification that jointly learns from image content and metadata while explicitly accounting for incomplete metadata and all the aforementioned challenges inherent in DICOM series image data.
%
Our main contributions are summarized as follows:
\begin{itemize}
    \item An end-to-end multimodal DICOM series classification framework that integrates visual and metadata representations using a bi-directional cross-modal attention (BCA) module to produce series-level representations with cross-modal and cross-slice contextualization.
    \item A sparse, missingness-aware metadata encoder (SME) that implements a learnable dictionary and FiLM to embed observed metadata index–value pairs. This encoding does not require any kind of missing value imputation, making it resilient to incomplete DICOM header data. 
    \item A flexible 2.5D visual encoder allowing each slice representation to attend to all other sampled slices, enabling the emphasis of relevant content while down-weighting redundant or irrelevant information.
    \item A comprehensive in-domain and out-of-domain evaluation on liver MRI series classification, demonstrating improvements over image-only, metadata-only, and multimodal baselines.
\end{itemize}

%

\section{Method}
\subsection{Model Architecture}

Our framework (Figure~\ref{fig:architecture}) treats DICOM series as a variable-length set of slices and associated metadata. From a series of $N$ slices, we subsample $S$ equidistant slices, yielding an image tensor $x \in \mathbb{R}^{S\times H \times W}$ and metadata tensor $y \in \mathbb{R}^{S\times F}$ where $H$ and $W$ are in-plane dimensions and $F$ is the number of metadata attributes. Each selected slice is center-cropped to $224\times224$, z-score normalized, encoded via an image backbone, and projected to a fixed dimension. To capture contextual dependencies between slices, we employ cross-slice attention \cite{vaswani_attention_2017} over slice-level tokens. Rather than aggregating slices independently, this mechanism allows each slice representation to attend to all other sampled slices, thereby enabling global contextualization. The final tensor containing the visual representations is denoted as $\mathbf{V} = [\mathbf{v}_1, \dots, \mathbf{v}_S] \in \mathbb{R}^{S \times d_v}$. In parallel, metadata is encoded with the Sparse Metadata Encoder (Section~\ref{method:sparse-metadata-encoder}) into fixed-size embeddings that represent observed DICOM metadata features $\mathbf{M} = [\mathbf{m}_1, \dots, \mathbf{m}_S] \in \mathbb{R}^{S \times d_m}$. Bi-directional Cross-modal Attention (Section~\ref{method:cross-attention}) then fuses all image and metadata embeddings, yielding cross-modal features that are then aggregated into a single series-level embedding, which is forwarded to the classification heads.

\subsection{Sparse Metadata Encoder (SME)}
\label{method:sparse-metadata-encoder}

Let $y \in \mathbb{R}^{S \times F}$ denote the metadata tensor for $S$ slices, and $F$ possible DICOM metadata attributes. Missing entries are represented as NaN and define a sparsity mask $\mathcal{O}_{s} \subset \{1,\dots,F\}$ of observed features for each slice $s$. Instead of treating metadata as a dense vector, we model it as a set of observed index-value pairs $\{(f, v_{s,f}) \mid f \in \mathcal{O}_{s}\}$.
Each feature index $f$ is associated with a learnable embedding $e_f \in \mathbb{R}^d$. To capture interactions between feature identity and its numeric value, we employ feature-wise linear modulation (FiLM)~\cite{perez_film_2017}. A value network $g_\theta : \mathbb{R}^{1+d} \rightarrow \mathbb{R}^{2d}$ predicts modulation parameters
$(\alpha_{s,f}, \beta_{s,f}) =  g_\theta \big( [\, v_{s,f}, e_f \,] \big)$, 
which produce a modulated embedding $\tilde{e}_{s,f} = e_f \odot (1 + \alpha_{s,f}) + \beta_{s,f}$. This contextualizes the scalar value $v_{s,f}$ by its semantic feature identity $f$. 
The modulated embeddings are aggregated across observed features using average pooling, yielding fixed-dimensional representation independent of the number of observed attributes, similar to set-based encoders \cite{zaheer_deep_2017}. Finally, the aggregated vector is refined via a residual MLP and projected yielding the final embedding $\mathbf{m}_s\in\mathbb{R}^{d_m}$. 

\subsection{Fusion with Bi-Directional Cross-Modal Attention (BCA)}
\label{method:cross-attention}

The vision and metadata embedding tensors $\mathbf{V}\in \mathbb{R}^{S \times d_v}$ and $\mathbf{M} \in \mathbb{R}^{S \times d_m}$ are linearly projected into a $d$-dimensional space, i.e. $\tilde{\mathbf{V}} = \mathbf{V} \mathbf{W}_v, \quad 
\tilde{\mathbf{M}} = \mathbf{M} \mathbf{W}_m$ with $\mathbf{W}_v \in \mathbb{R}^{d_v \times d}, \mathbf{W}_m \in \mathbb{R}^{d_m \times d}$. Cross-modal interactions are modeled via bi-directional multi-head attention (MHA) \cite{vaswani_attention_2017} with residual connections, layer normalization (LN), and a feed-forward (FF) network applied:
\begin{equation*}
\begin{split}
&\mathbf{V}' = \mathrm{MHA}(\mathbf{Q} = \tilde{\mathbf{V}}, \mathbf{K} = \tilde{\mathbf{M}}, \mathbf{V} = \tilde{\mathbf{M}}),
~~ \mathbf{V}'' = \mathrm{LN}(\mathbf{V}' + \tilde{\mathbf{V}} + \mathrm{FF}(\mathbf{V}' + \tilde{\mathbf{V}}))\\
&\mathbf{M}' = \mathrm{MHA}(\mathbf{Q} = \tilde{\mathbf{M}}, \mathbf{K} = \tilde{\mathbf{V}}, \mathbf{V} = \tilde{\mathbf{V}}),
~~ \mathbf{M}'' = \mathrm{LN}(\mathbf{M}' + \tilde{\mathbf{M}} + \mathrm{FF}(\mathbf{M}' + \tilde{\mathbf{M}}))
\end{split}
\end{equation*}
The modality-specific outputs are concatenated and projected with a learnable linear layer $\mathbf{W}_f \in \mathbb{R}^{2d \times d_o}$ to a $d_o$-dimensional output space:
\begin{equation*}
\mathbf{F} = \mathrm{GELU}(\mathrm{LN}([\mathbf{V}'' , \mathbf{M}''] \mathbf{W}_f)) \in \mathbb{R}^{S \times d_o}
\end{equation*}
Finally, a learnable MLP weighting function $w : \mathbb{R}^{S \times d_o} \to  [0,1]^{S}$ is used to aggregate slice-level embeddings into a single series-level representation:
\begin{equation*}
\mathbf{z} = \sum_{s=1}^S w(\mathbf{F})_s \mathbf{F}_s \in \mathbb{R}^{d_o}
\end{equation*}
This bi-directional design enables visual features and metadata to reciprocally modulate each other across slices.

\section{Experiments}
\subsection{Task and Datasets} For benchmarking, we use the public Duke Liver MRI dataset \cite{macdonald_duke_2023} and a large in-house cohort for the liver MRI series classification task. The Duke dataset contains 2,146 series with joint labels that combine sequence types, acquisition plane, and contrast phases. We merge closely related \textit{early arterial}, \textit{mid arterial}, and \textit{late arterial} into a unified \textit{Arterial} class, and merge \textit{late} and \textit{transitional} phases into a single \textit{Late} class, yielding 13 classes overall. Our in-house dataset comprises 82,134 de-identified series from multiple institutions and scanner vendors, with equivalent multi-label targets to those at Duke. The label space of both covers core sequence families (T1-weighted, T2-weighted), diffusion imaging and its derived maps (DWI, ADC), fat-sensitive techniques (Dixon in-phase/opposed-phase), MRCP, localizers, acquisition planes (axial, coronal, sagittal), and contrast-enhanced phases. 

\subsection{Evaluation}
We evaluate our approach in two tracks: (i) in-domain benchmarking on the Duke dataset via five-fold cross-validation, and (ii) out-of-domain generalization by training our model on the in-house cohort and testing on both its held-out partition and the Duke dataset. 
For in-domain benchmarking, we compare our approach to six baselines comprising uni- and multi-modal approaches, 2D/2.5D/3D inputs, and multiple metadata imputation and modality fusion techniques, see Table~\ref{tab:cv_setting}. All splits are class-stratified at patient level and use a unified preprocessing. For out-of-domain evaluation, Duke's joint labels are mapped to the in-house multi-label schema for compatibility. Performance is reported using the weighted F1 score across classes.

\setlength{\tabcolsep}{5pt}
\begin{table}[t]
\centering
\caption{Baselines for in-domain evaluation on the Duke dataset.}
\begin{tabular}{llllll}
\hline
Exp. & Modality& Image Enc. & MD Enc. &  Imputer & Fusion\\
\hline
(1) & Image & 2D DenseNet121 & N/A & N/A & N/A \\ 
(2) & Image \cite{zhu_3d_2022} & 3D ResNet18 & N/A & N/A & N/A  \\
(3) & Metadata & N/A & XGBoost & No & N/A \\
\hline
(4) & Joint & 2.5D DenseNet121  & MLP (dense) & Zero & Concat \\
(5) & Joint & 2.5D DenseNet121 & MLP (dense) & Learned & Concat \\
(6) & Joint \cite{miller_automated_2024}  & 2D DenseNet121 & RF & No & No\\
\hline
Ours & Joint & 2.5D DenseNet121 & SME & No & BCA \\
\hline
\end{tabular}
\label{tab:cv_setting}
\end{table}
\subsection{Implementation Details}
We selected the well-established DenseNet121 \cite{huang_densely_2017} as the backbone in the visual feature pathway. Our approach is not tied to this choice and other backbones are suited for medical imaging tasks \cite{truong_how_2021}. The slice number $S=10$ is selected via grid search (Table~\ref{tab:ablation}). The cross-slice attention is configured to produce tokens of dimension $d_v=256$. Metadata is encoded by the Sparse Metadata Encoder (SME) into embeddings of dimension $d_m=128$. After bi-directional cross-modal attention (BCA), joint features are aggregated into series-level embeddings of dimension $d_o=128$.
Training uses Adam optimizer with weight decay of $0.01$, a cosine learning rate schedule (base: $10^{-4}$, warm-up: $10\%$ of training steps) and gradient clipping between $[-0.5, 0.5]$. We train for 30 epochs in the in-domain cross-validation and 15 epochs in the out-of-domain setting, with batch size 64 on a single NVIDIA Tesla T4. 
The baseline models (2) and (6) are implemented according to the related original publications \cite{miller_automated_2024,zhu_3d_2022} and related git repositories. 
Concatenation baselines (4) and (5) use $S=3$ slices as inputs. (4) uses zero metadata imputation while (5) uses a small MLP to predict missing fields based on observed values and learnable baseline feature values.
We apply a consistent preprocessing scheme to the raw DICOM header values. Categorical tags are one-hot encoded, and a binary missingness indicator is appended per tag. This finally yields 119 metadata features.
\section{Results}
\subsection{In-Domain Evaluation}\label{sec:id-eval}
\begin{figure}[t]
    \centering
    \includegraphics[width=0.95\textwidth, trim={0.5cm 3cm 0.5cm 4.2cm}, clip]{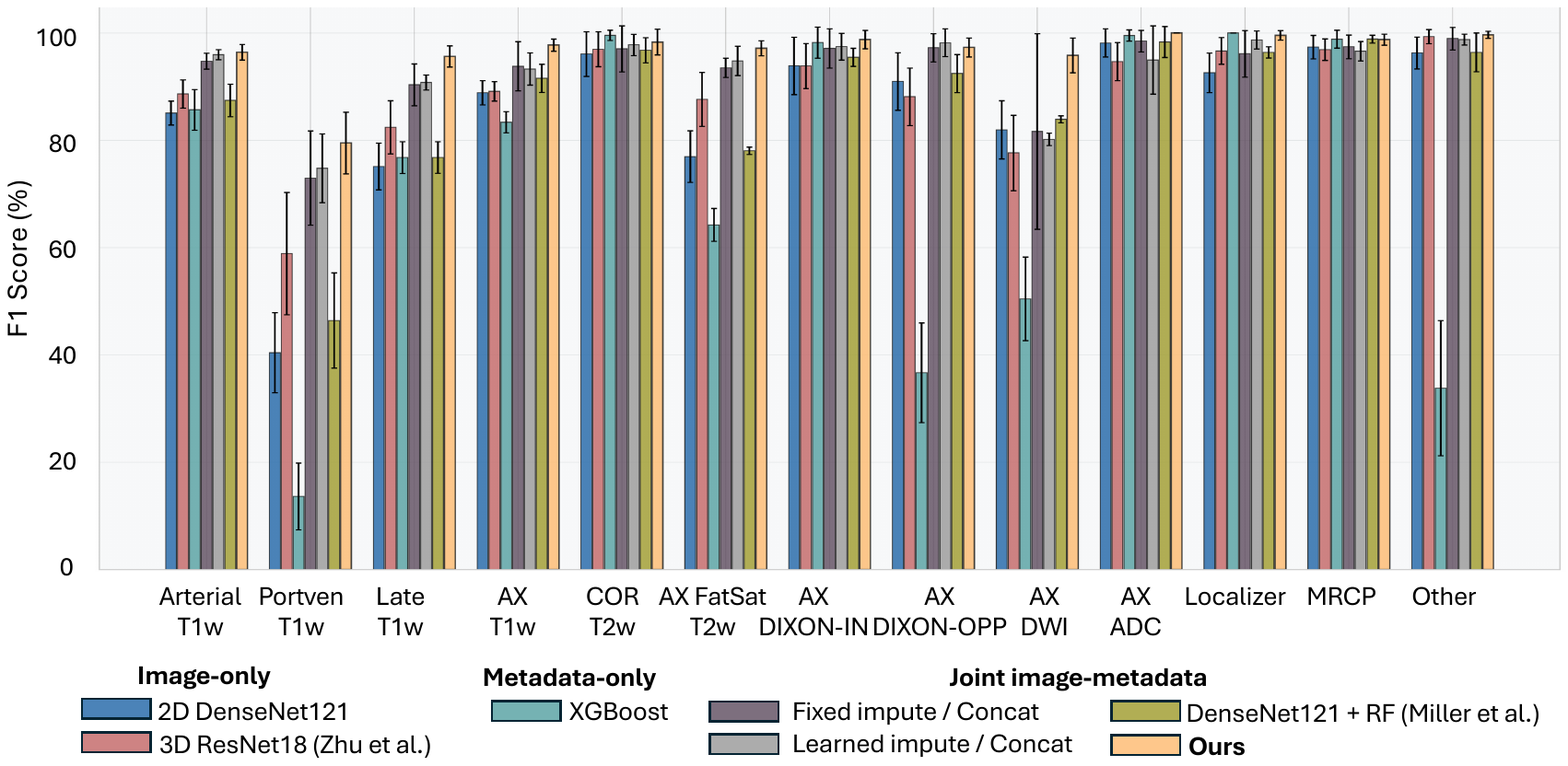}
    \caption{In-domain evaluation: five-fold cross-validation per-class F1 scores (\%) on the Duke Liver MRI dataset.}
    \label{fig:benchmarking-results-f1}
\end{figure}
The proposed method achieves the best performance across all evaluated methods (1)-(6) on the Duke Liver MRI dataset. In five-fold cross-validation (Table~\ref{tab:cv_summary}), our approach reaches a weighted F1 score of 96.66\% $\pm$ 1.03\%, outperforming all baselines with statistical significance ($p<0.05$, Wilcoxon signed-rank test).
The metadata-only baseline (3) achieves substantially lower performance (74.71 $\pm$ 2.34\%), indicating that acquisition metadata alone is insufficient for reliable series identification in this scenario. Per-class F1 scores (Figure~\ref{fig:benchmarking-results-f1}) indicate that the metadata-only approach can perform well for categories where related tags (e.g., acquisition plane) are directly available.
In comparison, image-only approaches (1) and (2) perform considerably better, demonstrating that visual information provides strong discriminative cues. Models (4) and (5) with end-to-end image–metadata fusion consistently outperform unimodal baselines, confirming the complementary nature of the two modalities.
Among multimodal approaches, the combination of SME and BCA provides advantages over simple imputation and feature concatenation techniques. The proposed strategy improves performance by approximately three percentage points compared to the best concatenation baseline (96.66\% vs. 93.51\%), highlighting the benefit of learned sparsity-aware metadata representations over dense metadata representation and learned modality interaction over static fusion. 

\begin{table}[t]
\centering
\caption{In-domain evaluation on Duke liver MRI sequence classification task. Baselines are (1) 2D Image-only, (2) 3D Image-only \cite{zhu_3d_2022}, (3) Metadata-only (XGBoost), (4) Joint: Concat + fixed imputation (zeros), (5) Joint: Concat + learned imputation (MLP), (6) Two-stage \cite{miller_automated_2024}. Our method with $S=10$.}
\begin{tabular}{lccc}
\hline
Exp. & Precision (\%) & Recall (\%) & F1 (\%) \\
\hline
(1)& 85.67 $\pm$ 1.22 & 85.37 $\pm$ 1.19 & 85.09 $\pm$ 1.31 \\ 
(2)& 88.77 $\pm$ 1.80 & 88.39 $\pm$ 1.89 & 88.33 $\pm$ 1.92 \\
(3) & 75.14 $\pm$ 2.39 & 76.66 $\pm$ 2.20 & 74.71 $\pm$ 2.34 \\
\hline
(4) & 93.83 $\pm$ 1.71 & 93.62 $\pm$ 1.76 & 93.51 $\pm$ 1.89 \\
(5) & 93.62 $\pm$ 3.02 & 93.37 $\pm$ 3.27 & 93.21 $\pm$ 3.48 \\
(6) & 87.81 $\pm$ 0.83 & 87.20 $\pm$ 1.16 & 87.01 $\pm$ 0.97 \\
\hline
Ours & \textbf{96.84 $\pm$ 0.98} & \textbf{96.67 $\pm$ 0.99} & \textbf{96.66 $\pm$ 1.03} \\
\hline
\end{tabular}
\label{tab:cv_summary}
\end{table}

\subsection{Out-of-Domain Evaluation}\label{sec:ood-eval}
%
As indicated in Table \ref{tab:out-of-domain-eval}, the performance on the in-house test split is high across all classes confirming the strong in-domain performance observed on the cross-validation experiments. The out-of-domain performance is measured by training our model on the in-house cohort and applying it to the Duke dataset. We observe that sequence-type classification remains strong for T2, DWI, ADC, and Dixon in-phase, with a moderate decrease for T1 and a larger drop for Dixon opposed-phase (74.07\% vs. 97.56\%). MRCP, localizer and acquisition-plane classification generalize well out-of-domain. For contrast phases, pre-contrast is similar (95.49\%), arterial is higher (92.16\% vs. 86.55\%), portal venous is lower (65.34\% vs 80.25\%), and the combined Late phase is comparable (88.21\%)  to in-domain late-phase performance. These particular differences may reflect concept shifts in protocol definitions or label schemas. 

\begin{table}[t]
\centering
\caption{ Out-of-domain evaluation: Model trained on in-house
cohort training split. Per-label F1 (\%) on the in-house holdout and the external Duke dataset.}
\label{tab:out-of-domain-eval}
\resizebox{\textwidth}{!}{
\begin{tabular}{llcc}
\hline
Task & Label & In-house test (in-domain) & Duke (out-of-domain)\\
\hline
\multirow{6}{*}{Sequence type}
& T1        & 97.77 & 92.27 \\
& T2        & 99.64 & 97.33 \\
& DWI       & 99.88 & 96.61 \\
& ADC       & 99.77 & 100.00 \\
& DIXON\_IN & 96.72 & 96.46 \\
& DIXON\_OPP& 97.56 & 74.07 \\
\hline
\multirow{2}{*}{MRCP}
& YES & 96.66 & 99.38 \\
& NO  & 99.83 & 99.95 \\
\hline
\multirow{2}{*}{Acquisition plane}
& AX  & 99.88 & 100.00 \\
& COR & 99.44 & 100.00 \\
\hline
\multirow{5}{*}{Contrast phase}
& PRE     & 97.98 & 95.49 \\
& ART     & 86.55 & 92.16 \\
& PORTENV & 80.25 & 65.34 \\
& TRANS   & 79.53 & \multirow{2}{*}{88.21} \\
& HEPA    & 88.58 \\
\hline
\multirow{2}{*}{Localizer}
& YES & 99.17 & 97.56 \\
& NO  & 99.90 & 99.88 \\
\hline
\end{tabular}
}
\end{table}
\section{Discussion and Conclusion}
We propose a method to address practical challenges of DICOM series classification in a unified way. The in-domain and out-of-domain performance of our method was quantitatively evaluated for Liver MRI series classification on two datasets and all considered baseline methods were surpassed.
The sparse, missingness-aware metadata encoder (SME) provides a simple mechanism to leverage multi-slice metadata without any kind of imputation. The lower performance of fixed or learnable imputation baselines (Table~\ref{tab:cv_summary}) suggests that imputation noise or complexity degrades classification performance. When missingness is high or training data is limited, imputation errors dampen gains from multimodal fusion.
%
The 2.5D strategy, combined with bi-directional cross-modal attention (BCA), integrates 3D image and cross-modal context while balancing computational efficiency. Using fixed-length slice sequences mitigates the unreliability of single-slice decisions while avoiding the complexity of full 3D modeling and naturally handles varying DICOM series lengths. The ablation related to varying slice counts (Table~\ref{tab:ablation}) confirms that cross-modal attention is benefiting from multiple tokens to align image and metadata representations and down-weight irrelevant information.
\setlength{\tabcolsep}{5pt}
\begin{table}[t]
    \centering
    \caption{Ablation on the number of input slices $S$. Weighted F1 scores (\%) on Duke (five-fold cross-validation) for $S \in \{1, 3, 5, 10, 20\}$ with all other settings fixed.}
    \begin{tabular}{ccccc}
        \hline
         \textbf{$S=1$} & \textbf{$S=3$} & \textbf{$S=5$} & \textbf{$S=10$} & \textbf{$S=20$}  \\
         \hline
         85.09 $\pm$ 1.79 & 95.46 $\pm$ 0.69 & 96.17 $\pm$ 1.00 & 96.66 $\pm$ 1.03 & 95.88 $\pm$ 0.82 \\ 
         \hline
    \end{tabular}
    \label{tab:ablation}
\end{table}

\textbf{Limitations}: The out-of-domain evaluation indicates that for a few specific categories (Dixon opposed-phase, portal venous), the performance is less robust suggesting that cross-institutional concept shifts  may remain a limiting factor for certain classes. In addition, the metadata contribution is modest for specific targets. This finding is consistent with prevalent missing or ambiguous header fields that reduce the utility of metadata and shift the discriminative burden to the image representations. Potential improvements include confidence-aware fusion, exploring advanced modulation rules beyond FiLM, richer parsing of protocol strings to enhance robustness under limited labels and heterogeneous metadata. Furthermore, we consider the application of the method to other medical imaging classification tasks as fruitful to further establish its effectiveness.
%


%
%
%
\bibliographystyle{splncs04}
\bibliography{references}

\end{document}